\documentclass[11pt]{article}
\usepackage{moriond,epsfig}

\def\be{\begin{equation}}
\def\ee{\end{equation}}
\def\bea{\begin{eqnarray}}
\def\eea{\end{eqnarray}}

\newcommand{\beq}{\begin{equation}}
\newcommand{\eeq}{\end{equation}}

\begin{document}
\vspace*{4cm}
\title{A Tale of Two Distances}

\author{Martin Kunz$^\dag$ and Bruce A. Bassett$^*$ }

\address{${}^\dag$ Astronomy Centre, University of Sussex, Brighton, BN1~9QH, UK\\
${}^*$ Department of Physics, Kyoto University, Kyoto, Japan  \&  
Institute of Cosmology and Gravitation, 
University of Portsmouth, Portsmouth, PO1~2EG, UK}

\maketitle\abstracts{
There are two basic ways to measure physical distances in cosmology: 
One based on standard candles and one based on standard rulers.
Comparing current data for each method allows us to rule out axion-photon
mixing and dust-extinction as the sources of supernova dimming and
generally protects the case for cosmic acceleration from attacks based on
loss of photons. The combined data constrains the energy densities
in a $\Lambda$CDM model to 
$0.19<\Omega_m<0.32$ and $0.47 < \Omega_\Lambda < 0.82$ (at $2\sigma$)
without recourse to any further data sets.
Future data will improve on these limits and allow us
to place constraints on more exotic physics.}

\section{Introduction}

The concept of absolute space and time disappeared in the transition
from the Newtonian theory of gravitation to General Relativity.
Nonetheless, there is a general duality in any metric theory of 
gravity implying that distances in cosmology are unique \cite{eth,ellis,SEF}.
The luminosity distance ($d_L(z)$, based on the apparent luminosity
of standard candles) and the angular-diameter distance ($d_A(z)$,
based on the apparent size of standard rulers) are linked by
distance-duality:
\beq
\frac{d_L(z)}{d_A(z)(1+z)^2} = 1\,.
\label{recip}
\eeq
where $z$ is redshift. 
Distance-duality holds for general metric theories of gravity in any background 
(not just Friedmann-Lema\^{\i}tre-Robertson-Walker [FLRW]) in which photons travel on 
unique null geodesics and is essentially 
equivalent to Liouville's theorem in kinetic theory. It is only valid if
photons are conserved, but is not violated by gravitational lensing
(for infinitesimal geodesic bundles).

In these proceedings we will start by comparing the two distances in the
standard FLRW framework. To show the power of distance-duality as a test
of non-standard physics, we then use it to rule out replenishing dust as
the source of the supernova dimming and to constrain axion oscillations
over cosmological distances. More details can be found in \cite{BK1,BK2}.

Our estimates of the luminosity distance $d_L(z)$ is provided by the latest
compilation of type-Ia supernova data \cite{riess}. This data set includes a 
significant number of $z>1$ observations. Our angular-diameter distance data, 
$d_A(z)$, come from FRIIb radio galaxies \cite{daly1,daly2}, compact 
radio sources \cite{crg,JD,jackson} and X-ray clusters \cite{allen}. It is important 
to remember that some of this data predated the discovery of acceleration by SN-Ia 
and that there are now completely independent, indirect, estimates of $d_A$, e.g. 
from analysis of the 2QZ quasar survey \cite{outram} (giving 
$\Omega_{\Lambda}=0.71^{+0.09}_{-0.17}$) and strong lensing from a combination of 
the CLASS and SDSS surveys with a maximum likelihood value of $\Omega_\Lambda = 0.74-0.78$ 
\cite{keeton}, in good agreement with estimates from radio sources.
The Sunyaev-Zel'dovich effect in galaxy clusters is a further possible
source of distance data \cite{reese,uzan}.

\section{Applications of distance-duality}

\subsection{Comparison within standard cosmology}

\begin{figure}
\center{\psfig{figure=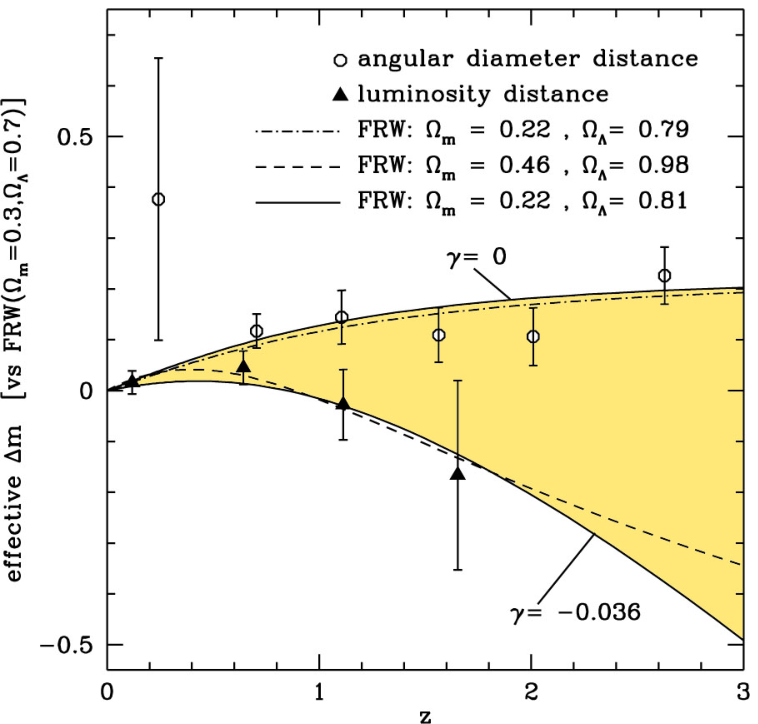,height=6cm} \qquad
	\psfig{figure=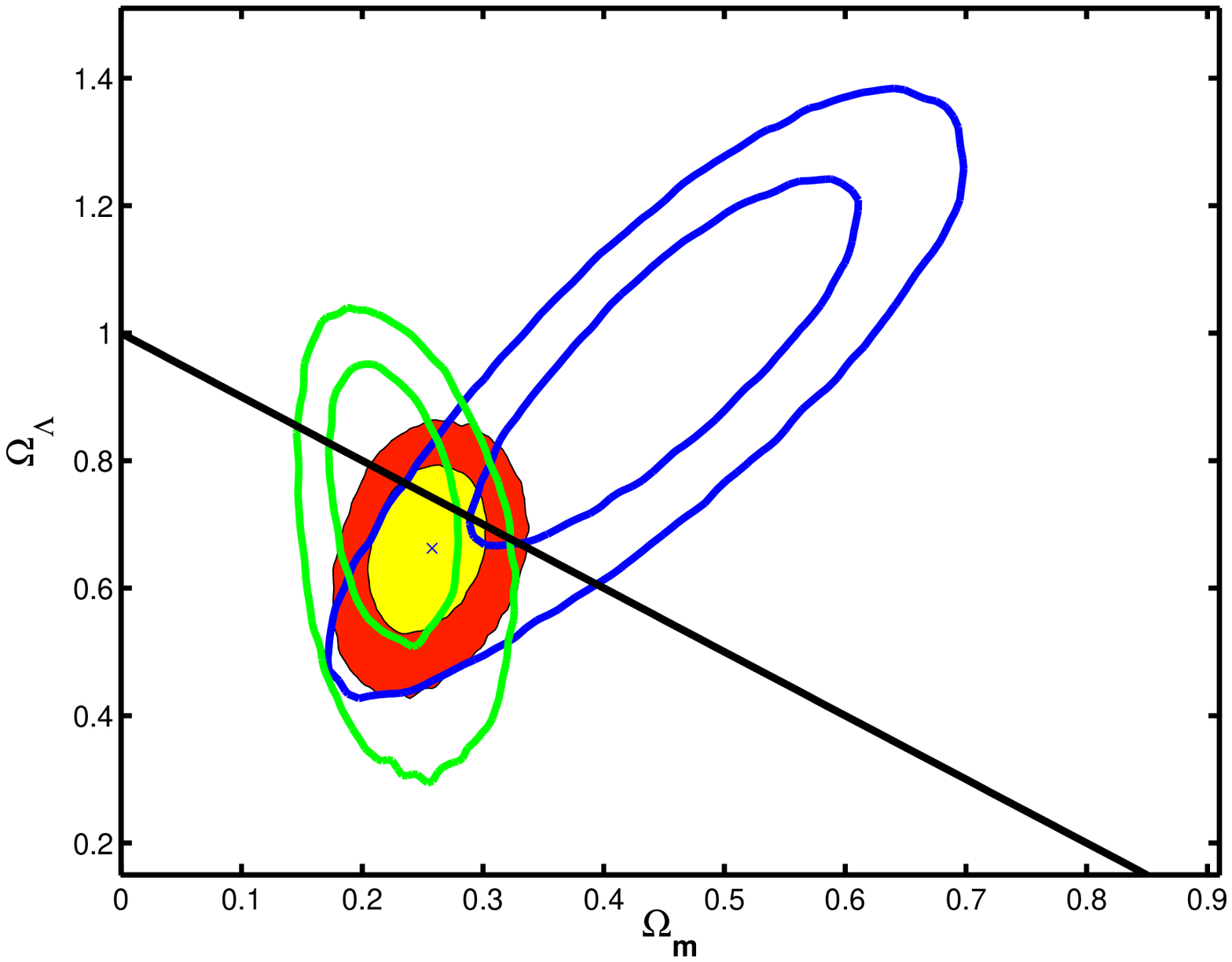,height=6cm}}
\caption{Left panel: The binned data for $d_L(z)$ (triangles, SN-Ia) 
and $d_A(z)$ (circles) 
are shown in equivalent magnitudes relative to the flat concordance model 
($\Omega_{\Lambda}=0.7,\Omega_m=0.3$) with $1\sigma$ error bars. They should 
coincide if distance-duality holds. 
The dashed curves are the best-fit FLRW models to the $d_A(z)$ data (top) and 
$d_L(z)$ (bottom) separately with no loss of photons. The solid 
curves have the same underlying FLRW model ($\Omega_\Lambda=0.81$, $\Omega_m=0.22$) 
but the lower curve includes the best-fit exponential brightening 
(see Bassett \& Kunz \protect\cite{BK1} for details). Right panel: 1 and 2$\sigma$
likelihood contours for $\Omega_m$ and $\Omega_\Lambda$ in a $\Lambda$CDM
framework for the angular diameter distance data alone (green/light-grey),
the luminosity distance data alone (blue/dark-grey) and the combined data
sets (filled contours).
\label{fig1}}
\end{figure}

As a first test, we plot the binned data as a function of redshift in
figure \ref{fig1}. We show equivalent magnitudes relative to the flat
concordance model ($\Omega_{\Lambda}=0.7,\Omega_m=0.3$) with $1\sigma$ 
error bars. Although the supernova data lies systematically below the
angular diameter distance data (and is thus {\em too bright}), the
violation of the distance duality is only at the $2\sigma$ level and
thus not significant enough to claim a detection \cite{BK1,uzan}.

Keeping this possible violation of the duality relation in mind, we
will nonetheless combine the two data sets to derive limits within
two frameworks: In the first one we assume a $\Lambda$CDM cosmology
while in the second one we restrict ourselves to flat universes, but
let the equation of state parameter $w=p/\rho$ of the dark energy
component vary (although we assume it to be constant). The right
hand panel of fig.~\ref{fig1} shows the $1$ and $2\sigma$ contours in
the $(\Omega_m,\Omega_\Lambda)$ plane for the SN-Ia data alone (blue/dark-grey
contours), the angular diameter data alone (green/light-grey contours) and the
combined data (filled contours). The diagonal black line shows the
flat models. The marginalised limits on the cosmological parameters
are given in table \ref{tab1}. In the second case, the supernovae
alone yield only very weak constraints on $w$ without additional
constraints on $\Omega_m$. The angular diameter distance data is less 
susceptible to this problem due to the larger redshift range, and
the combined data requires $-1.31 < w < -0.66$ at $2 \sigma$.

\begin{table}[t]
\caption{Limits on the cosmological parameters obtained by combining
current $d_L(z)$ and $d_A(z)$ data (the Hubble constant is always
being marginalised over).\label{tab1}}
\vspace{0.4cm}
\begin{center}
\begin{tabular}{|c|c|c||c|c|}
\hline
         & \multicolumn{2}{c||}{$\Lambda$CDM}& \multicolumn{2}{c|}{flat universe}       \\
data set & $\Omega_m$     & $\Omega_\Lambda$ & $\Omega_m$     & $w~(95\% {\rm CL})$     \\
\hline
$d_L(z)$ & $0.45\pm0.11$  & $0.94\pm0.19$    & $0.50\pm0.06$  & $-3.86^{+2.73}_{-6.36}$ \\
$d_A(z)$ & $0.23\pm0.04$  & $0.70\pm0.15$    & $0.22\pm0.05$  & $-1.00^{+0.40}_{-0.45}$ \\
combined & $0.25\pm0.03$  & $0.66\pm0.09$    & $0.25\pm0.05$  & $-0.94^{+0.28}_{-0.37}$ \\
\hline
\end{tabular}
\end{center}
\end{table}

Even the combined data sets are unable to constrain all three parameters
$(\Omega_m,\Omega_\Lambda,w)$ simultaneously, as universes with low
$\Omega_m$ and $\Omega_\Lambda$ together with a very negative equation
of state provide a good fit to all the data. Only the matter density
can be constrained in this case, $0.15 < \Omega_m < 0.31$ (95\% CL).

\subsection{Ruling out replenishing dust}

Riess {\em et al} \cite{riess} found that the best-fit model to all currently 
available SN-Ia was not an accelerating $\Lambda$CDM model but rather a replenishing 
grey-dust model \cite{goobar} with $\Lambda = 0$ which causes redshift-dependent 
dimming of the SN-Ia, with the evolution of $\rho_{\rm dust}$ changing from
$\propto (1+z)^3$ to a constant at $z=0.5$. If this 
was the correct explanation then we should expect a marked violation of distance 
duality with the $d_A$ data lying below the $d_L$ data since it would correspond 
to a non-accelerating universe. Our results show that this is not the case (indeed 
we have the opposite problem!)

A detailed analysis of this model based on \cite{goobar,ccm} gives a best-fit to 
{\em all} the data of 
$\Omega_{\Lambda} = 0.77 \pm 0.13$ showing that the combined data, in contrast to the 
SN-Ia data alone, rule out the replenishing dust model at over 4-$\sigma$.

\subsection{Limits on axion oscillations}

Another mechanism that was recently proposed \cite{AP} explains the supernova
dimming (relative to a $\Lambda=0$ cosmology) by allowing photons
to oscillate into axion states. In this way, about a third of the
photons are lost over cosmological distances. Again, as in the case
of dust, the angular diameter distance is unaffected, and should
thus correspond to the one expected for a standard CDM universe.
As fig.~\ref{fig1} shows, this is absolutely not the case, and
axion-photon mixing cannot explain away the need for a dark energy
component. 

We analysed this case in more detail \cite{BK2} by modeling the
transition probability as
\beq
P_{\gamma\rightarrow\gamma} = \frac{2}{3} + \frac{1}{3}e^{-l/l_{\rm dec}}
\eeq
and by introducing the dimensionless damping amplitude
$\lambda\equiv1/(2H_0 l_{\rm dec})$ which is zero if no
mixing occurs and one in the case of mixing over cosmological
distances. The combination of luminosity and angular diameter
distance data limits the mixing to $-0.7 < \lambda < 0.3$ and
the equation of state parameter of the dark energy component
to $-1.6 < w < -0.6$, both at $2 \sigma$.

A priori the absence of observed oscillations could be used to
place stringent constraints on the axion-photon coupling.
If we require the decay length $l_{\rm dec}$ to be of the
order of the Hubble scale and follow \cite{AP}, we end up
with an upper limit of $g_{a\gamma} \sim 1/M \sim 2\times10^{-12}/$GeV.
But this limit holds only for ultra-light axions, 
$m_a < 10^{-14}$ eV. The oscillation probability for
heavier axions is suppressed by a factor proportional to $1/m_a^4$.

\section{Discussion and conclusions}

We have shown that distance-duality is a powerful tool for 
constraining a variety of modifications of standard cosmology
as well as for improving our knowledge of cosmological
parameters. In particular, we are able to constrain the
energy densities in a $\Lambda$CDM universe at the 95\% confidence
level to $0.19<\Omega_m<0.32$ and $0.47 < \Omega_\Lambda < 0.82$
purely based on the expansion history of the universe. The
case $\Lambda=0$ is ruled out at very high confidence. 
This result does {\em not involve any perturbations} and is
thus not affected by issues like the initial spectrum of
perturbations or uncertainties in the determination of
$\Omega_b h^2$ and the reionisation optical depth.

We have further shown that there is no evidence for any strong
attenuation of high-redshift supernovae by dust (even dust
tailored to mimic the expansion-induced dimming) or for the
loss of photons due to axion-photon mixing.

With future experiments like the JDEM/SNAP satellite mission
\cite{SNAP} and the KAOS/gwfmos galaxy survey \cite{KAOS},
we expect to test deviations from distance duality at the level
of a few percent, implying that this diagnostic will mature
into a unique and powerful test of fundamental physics on
cosmological scales.

\section*{Acknowledgments}

We thank Sarah Bridle and Juan-Garcia Bellido for interesting
comments and questions, and Mark Hindmarsh for helpful explanations. 
MK is supported by PPARC. BB is supported
by the Royal Society and JSPS and thanks UCT for hospitality.

\end{document}